\documentclass[twocolumn,showpacs,preprintnumbers,amsmath,amssymb]{revtex4}
\usepackage{graphicx}
\usepackage{amsmath}
\usepackage{mathtools}

\newcommand{\ket}[1]{\mid #1\:\rangle}
\newcommand{\bk}[2]{\langle\: #1\mid #2\:\rangle}

\begin{document}

\title{Many worlds interpretation for double slit experiment}
\author{Zinkoo Yun}
 \email{semiro@uvic.ca}
\affiliation{Department of Physics and Astronomy University of Victoria, Canada}

\begin{abstract}
 As is well known, the double slit experiment contains every key concepts of quantum mechanics such as phase effect, probability wave, quantum interference, quantum superposition. In this article, I will clarify the meaning of quantum superposition in terms of phase effect between states. After applying standard quantum theory, it leads to serious questions about the unitary process of an isolated system. It implies that non collapsing interpretations including many worlds may not be justified. This also could explain that there is no such boundary between classical and quantum domains.
\end{abstract}



\maketitle

\section{Introduction}
\begin{figure}
\begin{center}
    \includegraphics[height=6cm]{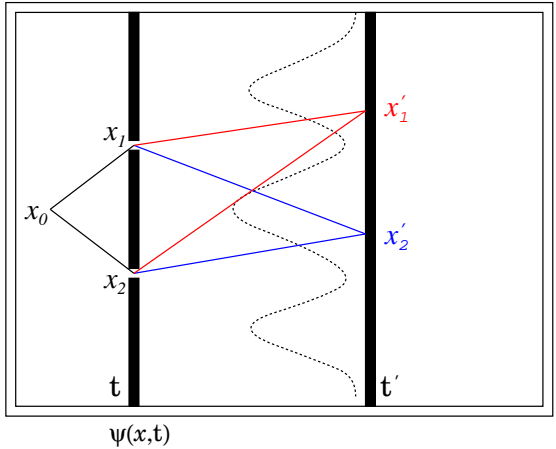}
  \end{center}
  \caption{ Double slit experiment in a box. Interference pattern appears as particles land at each spot $x'_i$ on the screen. Two position eigenstates $\ket{x_1}$ and $\ket{x_2}$ on slits are in quantum superposition, because there  is  non zero phase effect between them.}
\label{dslit_fig1}
\end{figure}

In early 1800s, Thomas Young has proposed an experiment which provides a strong evidence of wave theory of light.
Figure \ref{dslit_fig1} illustrates the experiment he proposed.
A stream of photons starts from $x_0$ and passes through two holes $x_1$ and $x_2$ on the slit and finally lands at spots on the screen. If the stream contains enough number of photons, we can observe interference pattern on the screen. 
 We have no problem of predicting exactly what this interference pattern looks like.
However, though more than 200 years have passed now since its discovery, it appears that we still do not completely understand this phenomena in quantum perspective, especially about the meaning of quantum collapse and measurement.

In this paper I will clarify about the meaning of quantum superposition in terms of quantum phase between states. Then after applying standard quantum mechanics, we will see what it implies to the unitary process of isolated quantum system and currently popular interpretations of quantum mechanics such as the many worlds interpretation.
\\

Since its first introduction by Hugh Everett\cite{everett, everett2} in 1957, and being ignored for quite a while before rediscovered by  Bryce DeWitt\cite{dewitt}, the theory of many worlds interpretation becomes popular and currently it is one of main interpretations of quantum mechanics accepted widely among many physicists because of its attractive feature of unitary evolution without collapsing wave function.

In order to see how the many worlds view interprets the double slit experiment, 
let's change the experiment setup little bit as shown in figure \ref{dslit_fig1}.  First, suppose we place a device at $x_0$ which emits a photon roughly every hour. This photon passes the slits and hits the screen eventually, where ``screen'' means the apparatus measuring position eigenstate of photon. Second, we put a stop watch on the screen, so that if the photon hits any position on the screen it records the time of the event. By this method, we can measure when the photon has landed on the screen. Finally let's put this whole system and apparatus in a box in order to make everything isolated.
After an hour, suppose we open the box to check the positions of photon landed on the screen. 

According to many worlds interpretation, the wave function of photon never collapses to one spot on the screen before opening box. We can express the state of system+apparatus+observer as a superposition of coexisting many worlds by
\begin{equation}\label{dslithw1b}
 \sum_i a_i \mid x'_i \:\rangle\mid A\, x'_i\:\rangle\mid OA\, x'_i\:\rangle
\end{equation}
for each spot $x'_i$ on the screen, where $\ket{x'_i}$ stands for the state of photon landed on the screen at $x'_i$.\footnote{We assume the photon is not destroyed by screen as we may perform the experiment using electron.} $\ket{A\, x'_i}$ stands for the states of particles of screen at $x'_i$ when the photon landed at $x_i$. $\ket{OA\, x'_i}$ represents the observer who observes the state of photon and the state of particles of screen after the photon landed at $x'_i$;
When we open the box, this many worlds split and we are subjected to one of them. Any non collapsing interpretation of quantum mechanics implies the entangled joint state similar to (\ref{dslithw1b}). Thus we cannot say that the photon has landed at any specific spot on the screen before we open the box. It has been in quantum superposition of all spots \emph{on the screen} before observation.

\section{The state of photon \emph{on screen}}
 In quantum mechanics, a superposition state means there is a phase effect between component states. In the example of double slit experiment in figure \ref{dslit_fig1}, we can measure it by observing interference effect between two quantum paths passing two slits. Thus we can say two position eigenstates $\ket{x_1}$ and $\ket{x_2}$ of photon on slit are in quantum superposition. 

Then, is the quantum state of photon \emph{on the screen} also in quantum superposition? In Feynman path integral technique  calculating the probability amplitude to measure the final position eigenstate  ($x_f, t_f$) from initial state $\psi(x,t)$, 
\begin{equation}\label{dslithwx1b1}
\langle x_f,t_f\mid \psi (x,t)\rangle
=\int dx \langle x_f,t_f\mid x,t\rangle \psi (x,t)
\end{equation}
we interfere all quantum paths ending up ($x_f, t_f$). We don't count any quantum path ending up other than ($x_f, t_f$) because they do not interfere with paths ending up  ($x_f,t_f$). Namely, two quantum paths ending up two different space time points do not interfere each other. That is, in figure \ref{dslit_fig1} experiment, any quantum path ending up at $x'_1$ and any quantum path ending up at $x'_2$ do not interfere each other. In other words, there is no  phase effect between  position eigenstates $\ket{x'_1}$ and $\ket{x'_2}$ of photon on the screen, so they are not in quantum superposition contrast to non collapsing interpretation.  That is, the state of photon on \emph{the screen} should be express by either $\ket{x'_1}$ \emph{or} $\ket{x'_2}$ not by their superposition $\ket{x'_1}$ \emph{and} $\ket{x'_2}$. Therefore even before opening box, the state of photon must be expressed by one of position eigenstate on the screen not by their quantum superposition like (\ref{dslithw1b}).

Of course, if there is no screen there, then  the quantum state of photon can be expressed by quantum superposition of position eigenstates in there, $\sum_i a_i \mid x_i\:\rangle$, because quantum paths of photon do not end up there, so we may measure interference effect between them.

\section{Meaning of superposition state} 
\begin{figure}
\begin{center}
    \includegraphics[height=4.5cm]{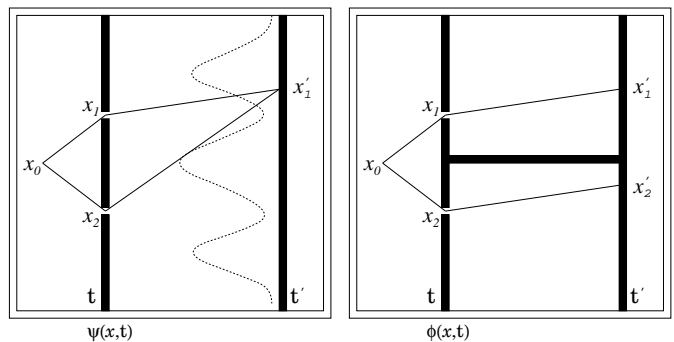}
  \end{center}
  \caption{(a) Two quantum paths end up at the same point $x'_1$.  They interfere, so two position eigenstates $\ket{x_1}$ and $\ket{x_2}$ are in quantum superposition; Two quantum paths did not end up at $\ket{x_1}$ and $\ket{x_2}$. They end up on the screen. (b)  Two quantum paths could not end up at the same point. They could not interfere, so $\ket{x_1}$ and $\ket{x_2}$ are not in quantum superposition.}
\label{dslit_fig2}
\end{figure}
What exactly does it mean by ``quantum superposition''? Let's remind the definition of this word.
Figure \ref{dslit_fig2}(a) describes typical double slit experiment and we put a divider in figure (b) experiment.
In figure  \ref{dslit_fig2}(a)  experiment, we can express the wave function of photon at slit as a quantum superposition of two position eigenstates $\ket{x_1}$ and $\ket{x_2}$,
\begin{equation}\label{dslithr1a}
\ket{\psi(x,t)}=a_1\ket{x_1}+a_2\ket{x_2} 
\end{equation} 
For given configuration of experiment, if the calculation of the relative phase effect between two states $\ket{\phi_1}$ and $\ket{\phi_2}$ could be non zero, then they are in quantum superposition. For example, suppose two quantum paths $\bk{f_1}{\phi_1}\bk{\phi_1}{ i_1}$ and $\bk{f_2}{\phi_2}\bk{\phi_2}{ i_2}$ pass two states $\ket{\phi_1}$ and $\ket{\phi_2}$. For arbitrary change of phase from $\mid\phi_1\rangle\langle\phi_1\mid$ to $e^{i\theta}\mid\phi_1\rangle\langle\phi_1\mid$ (from $\ket{\phi_1}$ to $e^{i\theta}\ket{\phi_1}$, if $\ket{\phi}$ is an initial or final state), if there exist quantum paths resulting in non zero change of (magnitude of) amplitude of measurement, we can say two quantum states $\ket{\phi_1}$ and $\ket{\phi_2}$ are in quantum superposition. If there is no quantum paths resulting in non zero change of amplitude by arbitrary phase change, then two states  $\ket{\phi_1}$ and $\ket{\phi_2}$  are not in quantum superposition.

In figure \ref{dslit_fig2}(a) experiment, if we change the phase along one path from $\mid x_1\rangle\langle x_1\mid$ to  $e^{i\theta}\mid x_1\rangle\langle x_1\mid$ arbitrary, the change of amplitude due to interference with another quantum path passing $\ket{x_2}$ could be non zero. Thus two quantum states $\ket{x_1}$ and $\ket{x_2}$ are in quantum superposition in figure  \ref{dslit_fig2}(a) experiment; However, the change of amplitude by arbitrary phase change is zero in figure \ref{dslit_fig2}(b) experiment. So $\ket{x_1}$ and $\ket{x_2}$ are not in quantum superposition in figure  \ref{dslit_fig2}(b) experiment. Thus the configuration of experiment could be a factor in determining whether two given states are in quantum superposition or not.

The reason is that because two quantum paths passing  $\ket{x_1}$ and $\ket{x_2}$ could end up at the same point in (a), so they could interfere each other, while they could not end up at the same point in (b). This reminds us the claim of Feynman path integral. Feynman path integral insists that if two quantum paths ends up at the same space time point, they interfere each other. Thus two states along each path could be in quantum superposition. If two quantum paths do not end up at the same space time point, they do not interfere each other. Thus two states along each path could not be in quantum superposition.
\\

With this standard in mind, let's consider whether two position eigenstates of photon \emph{on the screen} could be in quantum superposition in figure \ref{dslit_fig1} experiment.
The amplitudes of measuring $\ket{x'_1}$ and $\ket{x'_2}$ \emph{on the screen} are 
\begin{align}
\bk{x'_1}{\psi(x,t)}=a_1\bk{x'_1}{x_1}+a_2\bk{x'_1}{x_2} \label{dslithr2a}\\
\bk{x'_2}{\psi(x,t)}=a_1\bk{x'_2}{x_1}+a_2\bk{x'_2}{x_2} \label{dslithr2b}
\end{align}
For arbitrary change of phase from $\ket{x_1}$ to $e^{i\theta}\ket{x_1}$, the amplitudes do change. Thus $\ket{x_1}$ and $\ket{x_2}$ are in quantum superposition.

On the other hand, these amplitudes do not change as we change the phase of one quantum path arbitrary from $\ket{x'_1}$ to $e^{i\theta}\ket{x'_1}$ or from $\ket{x'_2}$ to $e^{i\theta}\ket{x'_2}$. Thus $\ket{x'_1}$ and $\ket{x'_2}$ are not in quantum superposition as expected from the fact they end up at different space time points.
Thus the phase difference between them does not mean anything physically. i.e., It is not measurable quantity in principle.

This fact does not matter how  photon interacts with screen. Someone may argue that 
\begin{equation}\label{dslithr2c}
  \parbox{0.8\columnwidth}{The photon landing at each position $x'_i$ interacts with the screen in complicate way.  If we count all this complications, then quantum state  $\ket{x'_1}\ket{A\, x'_1}$  and the quantum state  $\ket{x'_2}\ket{A\, x'_2}$  may show  phase effect between them  through the entangled state $\sum_i a_i \ket{x'_i}\ket{A\, x'_i}$.}
\end{equation}
It is possible to argue against (\ref{dslithr2c}) by
\begin{equation}\label{dslithr2d}
\parbox{0.8\columnwidth}{ All experimental data says that a photon physically interacts with  particles of screen at only one position. No data reveals that a photon physically interacts with particles of screen at multiple positions. Thus  we don't need to consider interference between particles of screen at multiple positions.}
\end{equation}
First of all, according to (\ref{dslithr2c}), two quantum paths in figure \ref{dslit_fig2}(b) also interfere each other, so that two eigen states  $\ket{x_1}$ and $\ket{x_2}$ are in quantum superposition; But most importantly, no matter which one is true, this is a null argument for the purpose of falsifying the other view:
(\ref{dslithr2c}) already assumes non collapsing of wave function and  (\ref{dslithr2d}) already assumes collapsing view. Regardless which one is true, we are not supposed to argue in this way to disprove collapsing or non collapsing view.
In disproving collapsing or non collapsing model, we are not allowed to use any claim which presumes collapsing or non collapsing view.\footnote{In this article the only basic assumption is that quantum paths of photon end up on the screen which both non collapsing and collapsing views would agree on. In some sense, the`` screen'' is defined in such a way.}
We have to argue by the criterion of definition of ``superposition'' itself.

In the analysis above, from the fact the quantum paths \emph{of photon} end up on screen, we have proved that $\ket{x'_1}$ and $\ket{x'_2}$ \emph{on the screen} are not in quantum superposition by the definition of it. In order for (\ref{dslithr2c}) to be true, $\ket{x'_1}$ and $\ket{x'_2}$ \emph{on the screen} must be in quantum superposition.

\begin{figure}
\begin{center}
    \includegraphics[height=6cm]{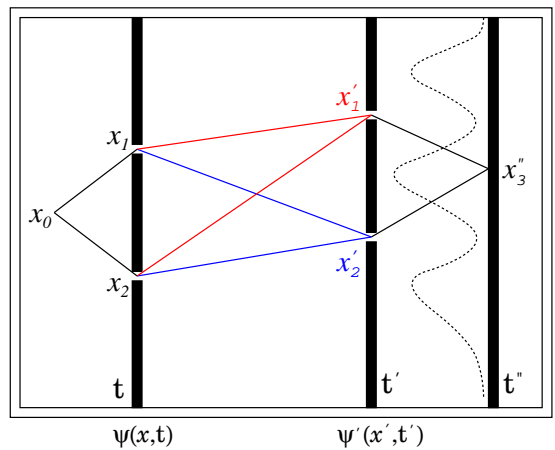}
  \end{center}
  \caption{After making two holes at $x'_1$ and $x'_2$ (or removing the first screen completely),  these two position eigenstates $\ket{x'_1}$ and $\ket{x'_2}$ become in quantum superposition. We can measure their phase difference effect on the second screen.}
\label{dslit_fig3}
\end{figure}
However  two position eigenstates $\ket{x'_1}$ and $\ket{x'_2}$ in there can be in quantum superposition, if the quantum paths do not end up there. Suppose we make two holes at $\ket{x'_1}$ and $\ket{x'_2}$ and put another screen at further distance as shown in figure \ref{dslit_fig3}. Then the amplitude of measuring the particle at $\ket{x''_3}$ is
\begin{equation}\label{dslithr2h}
\bk{x''_3}{\psi'(x',t')}=c_1\bk{x''_3}{x'_1}+c_2\bk{x''_3}{x'_2}
\end{equation}
Changing phase $\ket{x'_1}$ to $e^{i\theta}\ket{x'_1}$ arbitrary results in the change of amplitude. Thus in this configuration of experiment, two position eigenstates $\ket{x'_1}$ and $\ket{x'_2}$ are in quantum superposition. It implies that if we completely remove the first screen in figure \ref{dslit_fig3} experiment, the wave function in there is in quantum superposition of all position eigenstates $\ket{x'_i}$. This is expected because the quantum paths do not end up there. They end up at the second screen at further distance.

According to non collapsing interpretation of quantum mechanics including many worlds interpretation, in double slit experiment, before  observation (before opening box in figure \ref{dslit_fig1} experiment), the state of photon landed on screen is considered as a quantum superposition of all position eigenstates $\ket{x'_i}$ \emph{on the screen}. This superposition is accomplished through the entanglement $\sum_i a_i\mid x'_i\:\rangle\mid A\, x'_i\:\rangle$ between photon and screen which acts as a measuring apparatus. The analysis above proves that this cannot be true.
The analysis proves that two position eigenstates $\ket{x'_1}$ and $\ket{x'_2}$ \emph{on the screen} in figure \ref{dslit_fig1} experiment are not in quantum superposition. Thus before opening box,  the state inside box is one of $\mid x'_i \:\rangle\mid A\, x'_i\:\rangle$, not their quantum superposition. 
Thus we have proved that non collapsing  interpretation of quantum mechanics cannot be justified.

\section{Quantum entanglement and boundary}
We have proved that the state of photon in double slit experiment cannot be expressed by the entangled state between system and measuring apparatus like 
$\mid x'_i \:\rangle\mid A\, x'_i\:\rangle$. Let's discuss  its implication on boundary between quantum and classical world.

The state expressed by (\ref{dslithw1b}) is called the quantum entangled state. It is well accepted that if the  apparatus (in initial state $\mid A_0\rangle$) measuring eigenstates $\ket{S_i}$ of system is in contact with the quantum superposition state $\sum_i a_i\mid S_i\rangle$ of a system, then the resulting state before observation is expressed by the entanglement between system and the measuring apparatus,
\begin{equation}\label{dslithw1e}
\sum_i a_i \mid S_i\rangle\mid A_0\rangle \quad\to\quad \sum_i a_i\mid S_i\rangle\mid A_i\rangle
\end{equation}
where $\ket{A_i}$ stands for the state of apparatus measured $\ket{S_i}$ state of the system.
This entangled state is proposed in order to hold the unitary evolution of quantum theory. 
For example, (\ref{dslithw1e}) implies the quantum state  before observation, is  $\sum_i a_i\mid x'_i\:\rangle\mid A\, x'_i\:\rangle$ for photon and screen which we have disproved. Thus we know that generally the type of entanglement like  (\ref{dslithw1e}) is not possible. The entanglement between objects is possible only very special cases like decay of elementary particles.
Besides, we can see that the process (\ref{dslithw1e}) itself is a kind of non unitary process.

The entanglement process (\ref{dslithw1e})
between system and apparatus cannot be justified as we have demonstrated in double slit experiment.
\\

 ``Why the measuring device apply classical theory not quantum mechanics.'' This has been the unanswered question for a long history of quantum mechanics. Knowing that entanglement like (\ref{dslithw1e}) does not occur, we may answer this question; The truth is, in fact macroscopic apparatus do follow quantum mechanics exactly the same way as electron does. The difference is, as the mass increases, the dispersion of wave packet of Gaussian distribution becomes smaller according to Schrodinger's equation. For electron in static state, the dispersion speed is as the speed of light and for 1kg object the dispersion speed is 1m for hundred thousand years, so its dispersion can be ignored for our life time scale measurement. It means if we use 1kg object as an indicator of an experiment, we can almost trust our reading, but if we use a microscopic object as an indicator, we cannot trust our reading. That explains why Bohr insisted that measuring apparatus has to follow classical theory. In fact measuring apparatus also follows quantum mechanics, just its quantum effect is almost ignoble in our life time scale.  This shows that there is no boundary between quantum and classical realm.

\section{Conclusion} 
We have proved that non collapsing interpretation of quantum mechanics including many worlds cannot be compatible with standard quantum mechanics. In order to show it, the meaning of quantum superposition is clarified in terms of phase effect between states. Following the logic of standard quantum mechanics, it  leads to the conclusion  ``In general, the type of entanglement like (\ref{dslithw1e}) is impossible in principle.'' With this fact in mind, we could understand that there is no boundary between quantum and classical world.

\end{document}